\documentclass[sigconf,10pt,nonacm]{acmart}
\usepackage{xspace}
\newcommand{\sssec}[1]{ {{\flushleft \textbf{#1}}}}
\newcommand{\name}{SHARP\xspace}

\title{Privacy-Aware Sharing of Raw Spatial Sensor Data \\ for Cooperative Perception}

\author{Bangya Liu, Chengpo Yan, Chenghao Jiang, Suman Banerjee, Akarsh Prabhakara}
\affiliation{%
  \institution{University of Wisconsin-Madison}%
  \city{}
  \country{}%
}

\begin{abstract}
Cooperative perception between vehicles is poised to offer robust and reliable scene understanding. Recently, we are witnessing experimental systems research building testbeds that share \textit{raw spatial sensor data} for cooperative perception. While there has been a marked improvement in accuracies and is the natural way forward, we take a moment to consider the problems with such an approach for eventual adoption by automakers. In this paper, we first argue that new forms of privacy concerns arise and discourage stakeholders to share raw sensor data. Next, we present \name, a research framework to minimize privacy leakage and drive stakeholders towards the ambitious goal of raw data based cooperative perception. Finally, we discuss open questions for networked systems, mobile computing, perception researchers, industry and government in realizing our proposed framework.

\end{abstract}

\begin{document}
\pagestyle{plain}

\maketitle

\vspace*{-0.4cm}
\section{Introduction}
\label{sec:intro}
Cooperative perception is an emerging networked-systems application domain that involves sharing spatial sensor data between vehicles to enhance vehicle capabilities, safety, and user experience. As self-driving moves to higher autonomy levels, reliability is key. Cooperative perception, done correctly, can enable reliable machine vision detecting events ahead-of-time by extending the range of perception in both normal and adversarial weather. For example: An accident detected on a route can be used to reroute other vehicles. Black ice detected at a corner can be used to warn all following vehicles to turn on traction control. Sharing of such locally processed data ("events", bounding boxes, semantic labels) has been the focus of the initial wave of research and commercialization. However, \textit{raw spatial data} (camera, lidar and radar) sharing between vehicles is touted to bring the next wave of benefits from cooperative perception where the "whole is greater than the sum of its parts" is realized. For example: lidar return from small surface area objects like pedestrians yield only a handful of points, but when stitched with other vehicle's lidars, the densified point cloud can robustly detect critical objects. Using raw data as input to learning models has dramatically improved perception \cite{liu2023echoes, zhang2023robust,khan2024vrf,wang2020v2vnet,xu2022v2x}. We refer to data as "raw data" even though the actual communicated message is a compressed/encoded version, as long as it is not represented in a lossy style (like "events" or bounding box).


Realizing the power of raw spatial data demands innovation in perception, systems to realize high throughput, low latency, and methods to guarantee data confidentiality, integrity and availability. The expectations on all fronts are pushed to extreme levels than for processed data sharing. Recent works including RAO~\cite{zhang2023robust} and RECAP~\cite{recap} have extensively discussed the challenges within data synchronization and point cloud registration. While there are open challenges that are being addressed in computer vision, networking and mobile computing, we jump hoops to consider the eventual practical barrier: \textbf{privacy of raw data streams}. This problem is rarely considered by recent research, but we argue that it is of utmost importance for adoption by networked vehicles. We begin by systematically showing why this is a first-order concern. 



Today, auto manufacturers have privacy agreements with consumers that decide how data generated in their vehicle can be used. A single auto manufacturer can even share processed data within its ecosystem to generate timely safety alerts (e.g: Mercedes E \& S class). However, because of the diversity and the number of players in the auto industry, the vision of cooperative perception can be fully realized only when \textit{multi-automaker collaboration} is possible. An obvious challenge that emerges with multi-automakers is the mismatch in privacy agreements. One auto's consumer may opt-in for sharing data to third parties for added services, another may opt-out. Makers could also have different fine print in their agreements. So, how can the shared data be used? Should it be held to ego-vehicle (local) or collaborator's standards?  Beyond this, new forms of privacy concerns unique to raw spatial data emerge as follows: 
\sssec{Participating vehicle's location: } Typically, when spatial data is shared, for stitching purposes, accurate pose associated with the data frame composed of global location (from GPS) and a finer pose estimation (from local processing) is shared \cite{what2comm, where2comm, how2comm}. This pose data (used even with processed data sharing) would be shared under pseudonyms in line with privacy-preserving feature of vehicular communication standards \cite{henry2020pseudonym}. At first glance, this seems to reduce the utility of data to be short-lived and not linkable over time. However, just raw spatial data alone is rich enough to extract pose (through visual localization) \cite{naseer2018robust,toft2020long} and identify vehicles (through sensor noise, driving patterns, routes) over long-terms \cite{enev2016automobile,das2018deep}. Today's vision foundation models like VGGT \cite{wang2025vggt} and MapAnything \cite{keetha2025mapanything} make it easy to extract such data. The receiving vehicle could monetize these patterns with third-party insurance or advertising companies.

\sssec{Intellectual property of sensor design: } A huge concern for auto manufacturers is that when raw data is shared, the surface area for reverse engineering to learn trade secrets about low-level sensor quality is greatly exposed. Moreover, as mentioned above, despite pseudonyming, these low-level properties can be linked to automaker, auto type, sensor vendor etc. over long terms. Several billions of dollars are spent on research and development of spatial sensor hardware and integration. Beyond cost, sensors also vary greatly in quality. For example, one automaker may choose a camera vendor that offers superior stabilization as part of low level IP, another may have inferior stabilization. This begs the question: why should anyone participate in raw data sharing to give away secrets on the quality of their data?  



Concerns of such a high magnitude have made several automakers, auto consumers, and the public quite skeptical about raw data-based cooperative perception \cite{kpmg,ftc,privacy1,privacy2}. This paper presents \name \footnote{\name: \underline{SHA}ring \underline{R}aw spatial sensor data \underline{P}rivately }, a framework to alleviate privacy concerns and skepticism, and to encourage investment in all aspects (algorithms and systems infrastructure) relevant to cooperative perception with high fidelity, raw data.

Rather than dealing with these concerns at the cross-automaker privacy agreement level, the first question to ask is if past privacy-preserving computation frameworks (Secure Multi-Party computation and Fully Homomorphic encryption \cite{archer2023handbook}) are suitable to limit data sharing to specific functions. Although it is theoretically feasible to represent raw spatial data operations as additions and multiplications, the high number of samples in cameras, lidars and radars, and the communication overhead for decentralized operations render such options to be slow for real-time cooperative perception tasks with 10s-100s of milliseconds tolerances. 

\name's agenda is twofold. First, we propose a location obfuscation design that makes it challenging for raw data recipients to extract true auto location. Our design is inspired by the dramatic recent shift in 3D vision foundation models. Second, we look for inspiration in other industries where competitors have shared raw data for mutual benefit towards a shared goal. While safety can be a unifying cause (example: telcos should route all E911 calls, not just their customers), there are ways (like processed data sharing) to achieve simple metrics of safety without IP data leakage. To reap the benefits of raw data, we argue and lay out the need for a coordinated effort in defining open standards that preserve IP, multi-vendor agreements, and cost models.

\name demonstrates our line of reasoning, backed by a feasibility study, towards solving this major problem. Our hope is to enhance awareness of this seldom cared-for, last mile problem to multiple stakeholders and thereby spurring designs of privacy-cognizant holistic solutions. We also note that raw data sharing needn't be the de-facto standard for cooperative perception (with varying network capabilities), but should the infrastructure allow it and the environment demands it, it is imperative that we tackle this.

\begin{figure}
\centering
\includegraphics[width=\columnwidth]{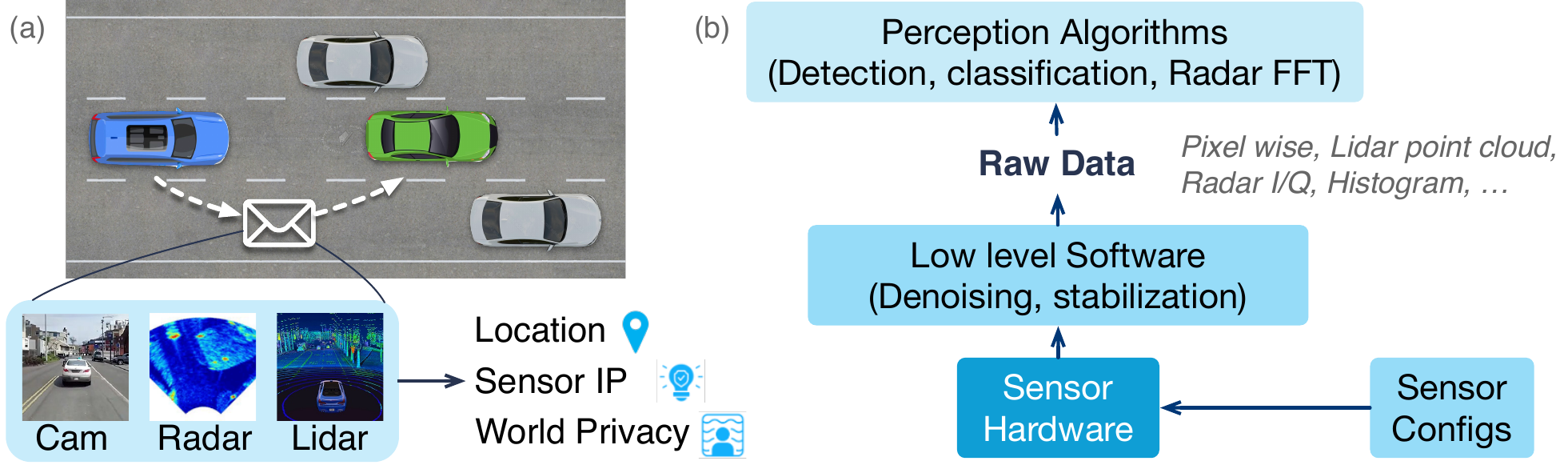}
\vspace*{-0.3in}
\caption{(a) shows the privacy leakage due to raw data streaming from blue car to the green car; (b) abstracts typical sensor architectures and shows raw data types}
\vspace*{-0.2in}
\label{fig:1}
\end{figure}

\section{The Need for Raw Data}
\label{sec:raw}
Raw spatial sensors surpass other low bandwidth data types in cooperative perception. Let us look at advantages that only raw data sharing can offer. 

\sssec{Cameras: } Visible light cameras are sensitive to lighting and weather, and
natively lack depth understanding. Cooperative perception can compensate for these by sharing (1) specific objects-of-interest (e.g: accidents) that are already detected locally by algorithms running on pixel-wise data; (2) pixel-wise image data to yield enhanced 3D understanding. 
Structure-from-motion, photogrammetry, neural 3D understanding all leverage pixel-wise data \cite{mikhail2001introduction,fu2024colmap,sumikura2019openvslam,mildenhall2021nerf,colmap}, which offers higher fidelity perception and is widely used.

\sssec{Lidars: } Lidars are sparser than cameras. Conventionally, the rawest level of data made available by vendors is point clouds obtained from the first return from specific laser directions. So, the lightest way to share lidar data is bounding boxes on 3D point cloud, ETSI standards \cite{etsi} for example. 
However, for weak objects or cars in foggy conditions where the range is reduced, detection is hard with only a few spatial points. With point cloud sharing or full time delay histograms available with SPAD lidars, we can obtain dense aggregated point clouds and thus, robust object detections. 

\sssec{Radars: } Solid-state radars are orders of magnitude sparser than cameras and lidars. Typical processing involves thresholding of raw histogram data to convert to sparse point clouds. 
The first level is the accumulation of point clouds that leads to the densification of objects and better object detection \cite{bansal2020pointillism}. Second, we have exchange of intensity-only raw data. 
Third, phase-sync aggregation (aka coherent aggregation) can dramatically boost the detectability of weak objects such as a pedestrian standing next to a bright object like a stop sign. 
Here, we consider I/Q data and unthresholded, intensity-only data as "raw radar data".

We note that low-level signal processing blocks (see Fig \ref{fig:1}) might be attached to sensors and only then expose raw data. What we refer to as "raw data" also varies from sensor-to-sensor. It is clear that there is a data hierarchy. Raw data is at the top of this hierarchy, offering the highest fidelity and enabling algorithms that enhance the degree of reliability.

\section{\name's Design}
\label{sec:design}

In this section, we alleviate privacy concerns in Sec. \ref{sec:intro}.

\subsection{Tackling vehicle's location leakage}

\label{ssec:loc_leakage}
\begin{figure}
\centering
\includegraphics[width=\columnwidth]{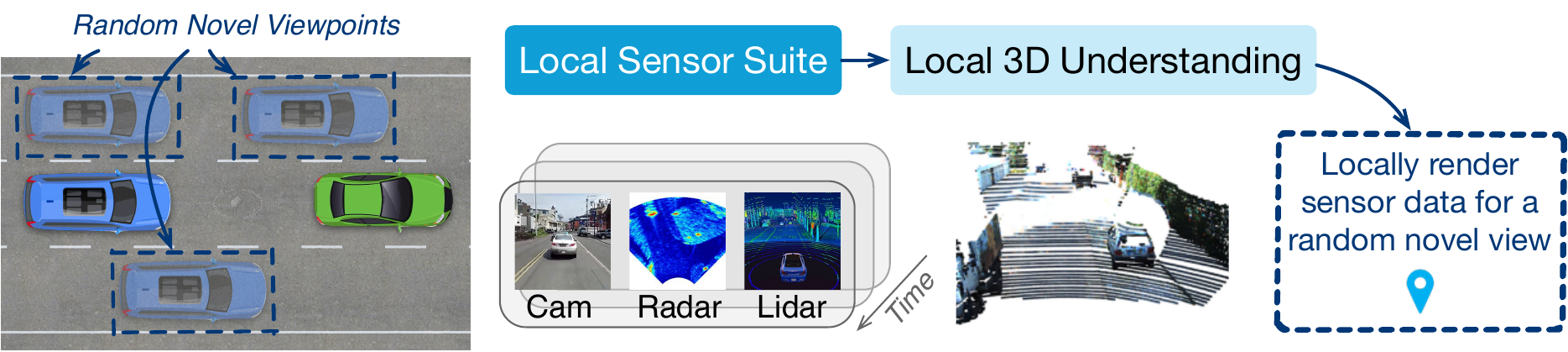}
\vspace*{-0.3in}
\caption{\name's approach to hide true location: use \textit{fast} 3D understanding and rendering to constantly spoof realistic, raw data from a random view point.}
\vspace*{-0.3in}
\label{fig:2}
\end{figure}

Sharing raw spatial streams (cameras, lidars, radars) inevitably enables recipients to reconstruct sensor and vehicle poses. Our key insight: generating sensor measurements from alternate viewpoints protects the true location (Fig \ref{fig:2}). By constantly randomizing viewpoints, true location tracks remain hidden even when pseudonyms are constant. 

\begin{figure}
\centering
\includegraphics[width=0.7\columnwidth ]{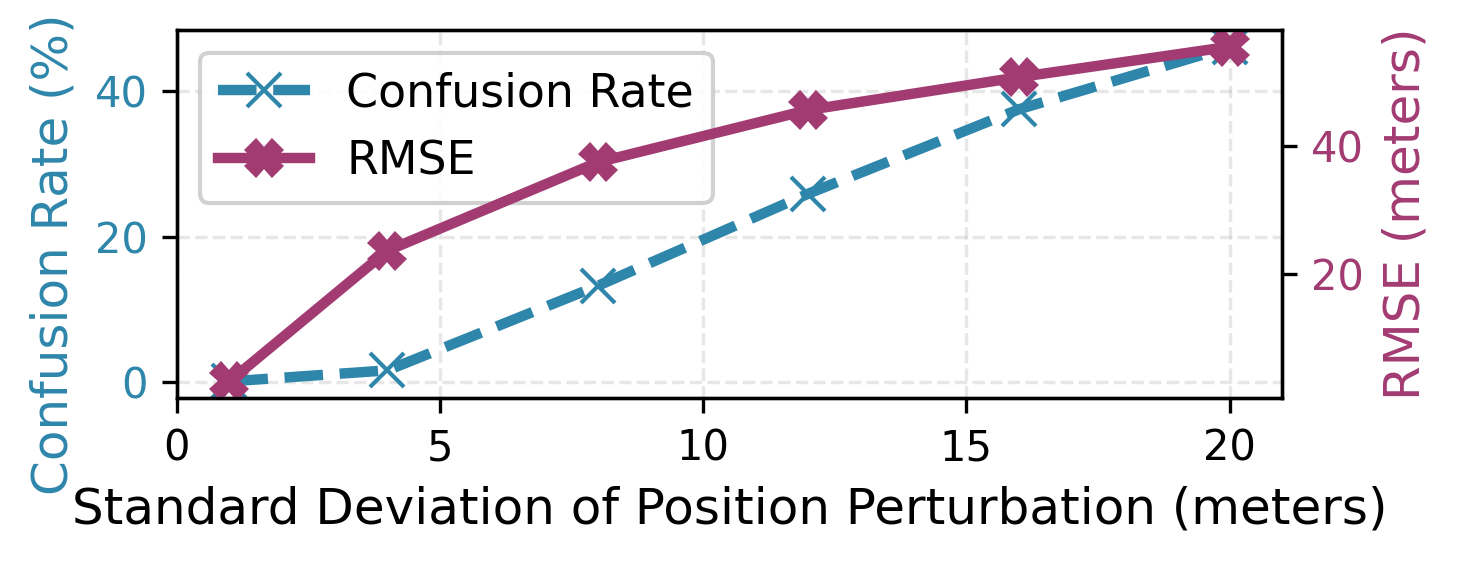}
\vspace*{-0.2in}
\caption{Preliminary simulation on the OPV2V scenes.}
\vspace*{-0.2in}
\label{fig:largescalesim}
\end{figure}

\sssec{Preliminary simulation:} To quantify the feasibility of this privacy paradigm, we conduct a large-scale simulation across all 73 scenes in the OPV2V \cite{xu2022opencood} dataset, with 50 rollouts per scene. For each rollout, we randomly select one vehicle as the ego vehicle (i.e., the receiver) while surrounding vehicles (i.e., sharers) share perception data at forged locations—coordinates perturbed by Gaussian noise. We then evaluate the privacy level of these sharers from two perspectives: (1) the deviation of forged trajectories from ground truth trajectories, and (2) the likelihood that the ego vehicle can correlate a sharer's identity with a physical vehicle observed in the environment. Within the ego vehicle, we deploy a nearest neighbor algorithm for tracking sharers and employ the Hungarian algorithm for matching between inferred and physical trajectories. We use RMSE and N-to-N matching error rate (confusion rate) as evaluation metrics, with results presented in Fig.\ref{fig:largescalesim}. The results reveal a clear trend: as the offset between forged and ground truth positions increases, the ego vehicle's ability to identify sharers diminishes. Specifically, at an offset of 12 meters (equivalent to 4 lanes of line-of-sight distance), the confusion rate reaches 25\% and the overall RMSE exceeds 45 meters. This large-scale simulation demonstrates \name's effectiveness in protecting sharer privacy.

\sssec{Challenges:} Although obfuscation can be achieved, such a solution is valid only if the sensor measurements generated are realistic. Novel View Synthesis (NVS) involves 3D reconstruction from local sensor measurements and rendering novel views. Approaches range from classical photogrammetry \cite{mikhail2001introduction} and SLAM \cite{sumikura2019openvslam} to recent neural methods \cite{mildenhall2021nerf,fu2024colmap}. However, 3D reconstruction takes seconds to hours—incompatible with automotive perception at 10+ Hz. Context-specific spoofing \cite{shenoy2022rf} lacks generalizability, necessitating full 3D understanding.

\sssec{Our approach:} Fortunately, recent breakthroughs like Dust3r \cite{dust3r} and Visual Geometry Grounded Transformers (VGGT) \cite{wang2025vggt} have revolutionized 3D understanding, reducing reconstruction time to milliseconds—orders of magnitude faster than prior work. Rather than iterative optimization, large-scale feed-forward models perform fast inference from 2D images to 3D points, similar to monocular depth estimation \cite{godard2019digging} but with dramatically improved quality through massive training data and model scale. This paradigm shift enables low-latency automotive applications. We propose vehicles to use VGGT-like models \cite{wang2025vggt} for local 3D understanding, and then render novel views at randomized locations in real time. From our simulation, such an obfuscation creates confusion in extracting the true vehicle pose. Sec \ref{sec:performance} shows the feasibility and realism of the novel views. VGGT requires extensions for comprehensive automotive deployment. Sec \ref{sec:disc} discusses open challenges, including occlusion-aware viewpoint selection.

While novel view based obfuscation masks exact trajectories they can still reveal general location (e.g., an intersection). To prevent receivers from building long term associations of general locations, we need to make it more challenging to map time-varying pseudonyms to uniquely identifiable sensor traits.  Sec \ref{ssec:ip} masks low-level sensor details and prevents building long term location tracks based on general locations.







\subsection{Tackling IP leakage}
\label{ssec:ip}

IP leakage from sharing raw spatial sensor data is so serious a concern that it could just render all research efforts towards raw data based cooperative perception moot. 

\sssec{Tiered relationship in automotive markets: } In experimental testbeds built with development kits \cite{zhang2023robust}, one can access raw data. However, the auto industry is layered as tier 1, 2, \& 3 depending on the level of interaction with the final automaker. The vast majority of final automakers only have access to processed sensor data from tiered vendors due to IP protection. Vertical integrators like Waymo and Tesla, either make sensors in-house or have special contracts to get hardware from non-competitors with access to raw data (Samsung, LG) for building their own proprietary perception. In a competitive scenario, why should one with access to raw data, share and leak secrets? Well, Sec \ref{sec:raw} has shown significant benefits, so this is a utility-privacy trade-off.

\sssec{Strawman solutions:} Approaches like neural encoders are insufficient, because to use raw data, a trained decoder \cite{wang2020v2vnet} at the receiver is needed and one would then have low-level sensor access. We could also consider cryptographic one way functions. But, they are not designed with the intention to preserve spatial correlations. So, the receiver would need to convert back to raw data before running cooperative perception. Another solution could be to add noise to the raw data to make it confusing for IP theft. However, each hardware has different capabilities (sampling rate, pixel / antenna count, sensitivity etc.), and metadata about these capabilities should be shared for the recipient to fully use the raw data. This would directly give away critical data.

\sssec{Our approach: }We call for a common unifying representation that all automakers (with raw data) can agree on. To this end, we argue that a open low-level stack from just above the sensor hardware till "raw data" must be executed to arrive at a common representation. We envision that such a stack would be built using public knowledge. It is quite possible that such a stack will be inferior to running a proprietary stack. But atleast, such data can be shared without any IP concerns. Thus, we propose a common stack for the purpose of raw data cooperative perception.


\begin{figure}
\centering
\includegraphics[width=\columnwidth]{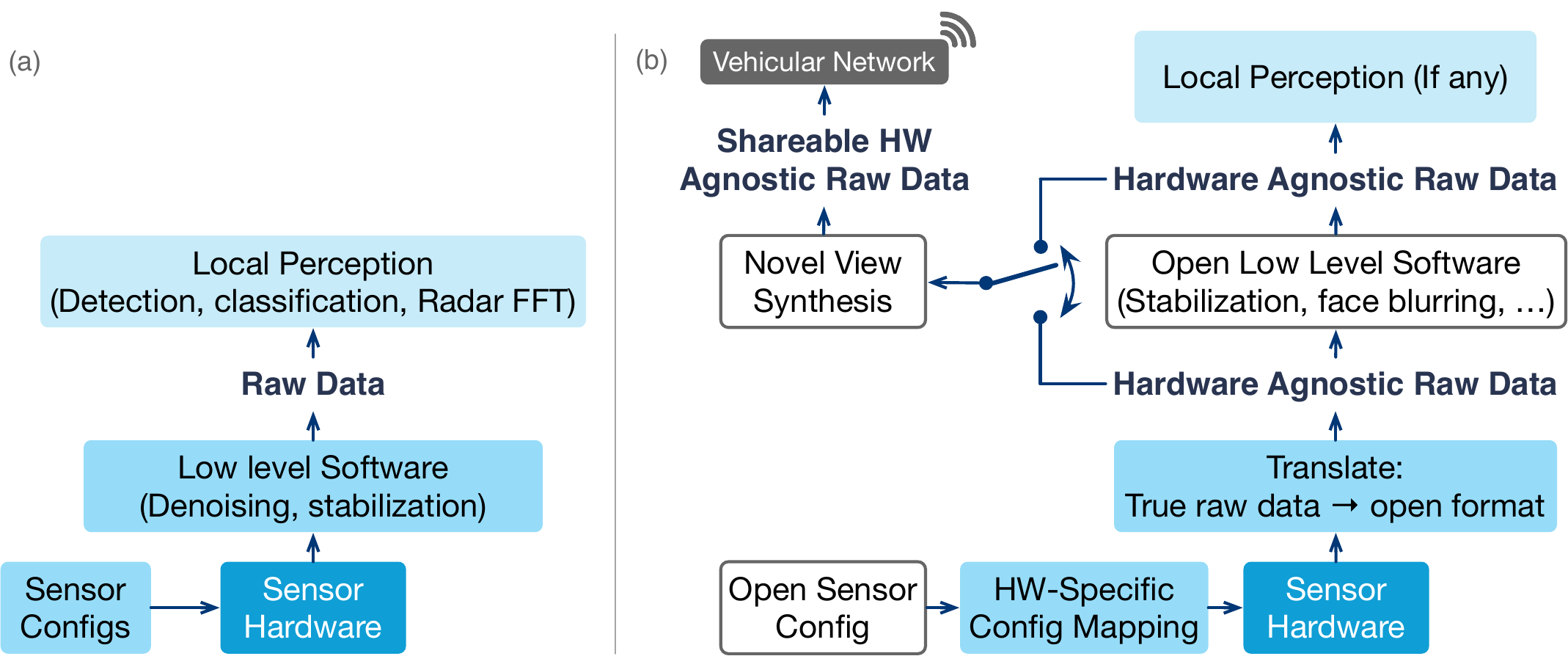}
\vspace*{-0.2in}
\caption{\name's approach to minimize IP leakage: Open sensor configurations, open low-level software and standardizing hardware agonstic raw data as output from sensor hardware. Blue blocks are proprietary. We schedule (b) to create shareable raw data, else we schedule a fully proprietary stack (a).}
\vspace*{-0.25in}
\label{fig:3}
\end{figure}
To fully execute an open standard, one needs not only an open low-level signal processing block but also the ability to control sensor configs and hardware. For example, even though a vehicle may have 10 antennas, the open config could expose only 1 antenna's data -- protecting the vehicle's secret about number of antennas. Standardizing hardware is simply infeasible! Modifying sensor configs and running a low-level software block are also inconvenient, because each vehicle may want to run their proprietary stack regardless -- for local perception.

Our idea is to offset the inconvenience by carefully swapping from proprietary stack (Fig \ref{fig:3}a)  to the open stack (Fig \ref{fig:3}b) at a duty cycle. In designing a suitable scheduling algorithm, the key constraints to consider are computing power, latency arising as a result of swaps, and the overall end-to-end latency needed for cooperative perception reaction times in challenging environments (bad weather, etc.). For example, one can switch to open configs and run open stack every once in 200 ms (5Hz) or based on demand, whereas the proprietary stack runs every 50 ms (20 Hz). 


\sssec{Standardization: } To implement this, we still need to deal with hardware IP. Hardware varies widely in capabilities, and sensor vendors often tightly integrate hardware, configs and low level software. While configs and low-level software can rely on public, open stacks, we need an interface to various hardware. Thus, we have arrived at a bottleneck which can only be resolved via standardization. We envision bringing different hardware sensor manufacturers together to implement a \textit{compact proprietary layer} to map their true hardware output to an agreed upon, \textit{hardware-agnostic, open format}. On top of this, other features can also be implemented on an open stack (see Fig \ref{fig:3}). For example: even if a radar uses a high sampling rate, a sensor vendor could modify the data to only satisfy higher-level open format requirements, on radar range and resolution. Other hardware variations to be addressed in the proprietary layer include waveform parameters, effect of sensor noise, mosaicing, size, field of view, gain, number of lidar beams, number of antennas etc. Similarly, the manufacturer would also need an interface to map the open sensor configurations to the hardware capabilities. This calls for efforts to (1) create a taxonomy and represent high-level hardware capabilities and define shareable open formats, (2) build (proprietary) software to translate open configs to native hardware, (3) build (proprietary) software to translate true raw data to open format.

\sssec{Cost models: } Despite the need to share raw data for mutual safety, a stack swapping technology solution is likely to face some opposition (for resource-constraints reasons), unless we incentivize it as a revenue-generating stream. We argue for a cost model that a recipient pays for if they subscribe to raw data. Since the stack is open and based on publicly available information, we can treat each data to be of identical value. 
The total cost would be based on how frequently the recipient asks data, and any priority levels that they expect to be served at. For example: in harsh weather or when a vehicle is behind a long truck on a single lane highway, that is, when raw data need is extreme, we expect to switch priority to higher levels. We expect multi-party agreements facilitated through a lightweight, common billing system.

By scheduling stack swapping, creating standards for an open stack and incentivizing automakers with additional revenue stream, \name\ facilitates raw data exchange. As we build this, we should also work towards integrating it with default automotive embedded workflows.


\section{Feasibility}
\label{sec:performance}
Here, we show the privacy problem with raw data and feasibility of our approach. We run experiments on OPV2V \cite{xu2022opencood} driven by Carla simulator \cite{carla} with Nvidia V100S 32GB.


\sssec{Vehicle pose could be easily inferred from raw data: } We perform a demonstrative experiment that estimates the camera poses from streams of unconstrained RGB images. 183 streams with 10 frames each. VGGT as a vision foundation model can predict camera poses and depth maps simultaneously. Table. \ref{tab:vscolmap} shows the fast inference, and accurate pose from VGGT compared to traditional method like COLMAP \cite{colmap}  as well as the SoTA visual SLAM work, DROID-SLAM \cite{droid}, confirming that location privacy is a serious concern.

\begin{table}
    \centering
    \small
    \begin{tabular}{|c|c|c|c|}\hline
         Method&  success rate&  pose err(rmse)&  time(s)/frame\\\hline
         VGGT&  183/183&  0.6900&  0.089\\\hline
         DROID&  183/183&  1.2982&  0.166\\ \hline
 COLMAP& 43/183& 2.8477&5.150\\\hline
    \end{tabular}
    \caption{A receiving car can easily estimate camera pose from received raw images with VGGT.}
    \vspace*{-0.3in}
    \label{tab:vscolmap}
\end{table}

\begin{figure}
\centering
\includegraphics[width=0.8\columnwidth ]{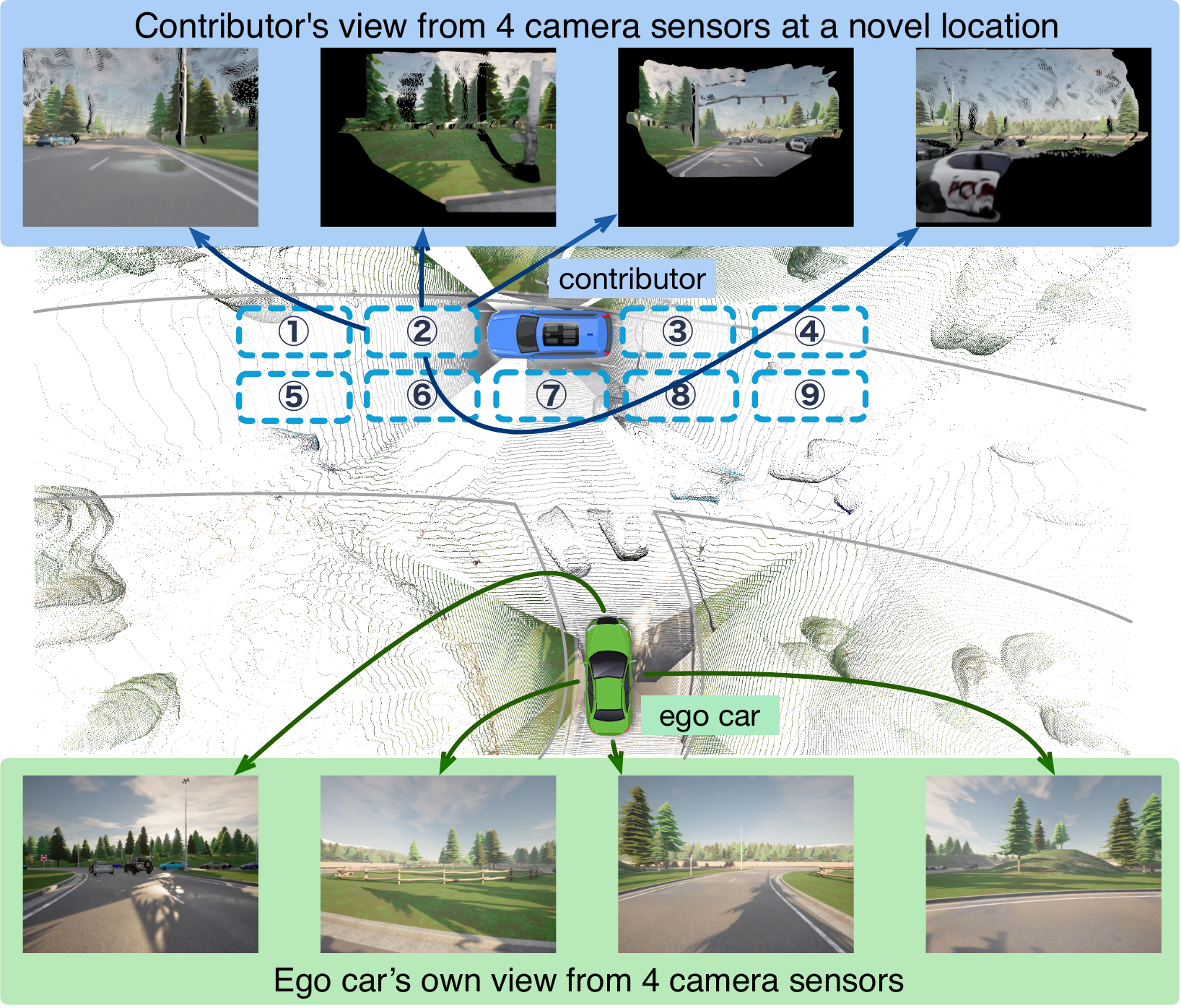}
\vspace*{-0.15in}
\caption{A demo on Town06 of OPV2V dataset, where contributor pretends it is at location 2.}
\vspace*{-0.15in}
\label{fig:simple_demo}
\end{figure}

\sssec{\name's novel view synthesis to obfuscate vehicle location is effective: } As proposed in Sec \ref{ssec:loc_leakage}, contributor performs 3D scene understanding with local data via vision foundation model, then renders sensor data from a novel view, and only such novel view is shared to an ego-agent. The rendering could be achieved by either plain point projection or through 3DGS \cite{3dgs} to reduce pixel holes. The ego agent then leverages the received views and its own sensor data to understand the surrounding. 
Fig \ref{fig:simple_demo} shows a qualitative result. Table \ref{tab:error_map} shows the overall relative depth estimation error \footnote{Output of VGGT is non-metric, hence popular benchmark on object detection is not a suitable evaluation task in this case.}. Ego car achieves a lower error with cooperative perception, benefiting from shared data that is either true or from a novel view. Across all 9 locations, the error is comparable and generally closer locations yield lower errors. 

\begin{table}
    \centering
    \small
    \begin{tabular}{|c|c|c|c|c|c|}\hline
         \textbf{Shared Data}&  @1&  @2&  true &  @3&  @4\\\hline
         \textbf{MSE} &   0.0108&  0.0083& 0.0088&  0.0089&  0.0117\\ \hline
 No Coop& @5& @6& @7& @8& @9\\\hline
 0.0142& 0.0127& 0.0101& 0.0086& 0.0117& 0.0103\\\hline
    \end{tabular}
    \caption{Relative depth error when ego agent does no cooperative perception and when sensor data from different viewpoints are shared by contributor.}
    \label{tab:error_map}
    \vspace*{-0.4in}
\end{table}

\begin{figure}
\centering
\includegraphics[width=0.95\columnwidth ]{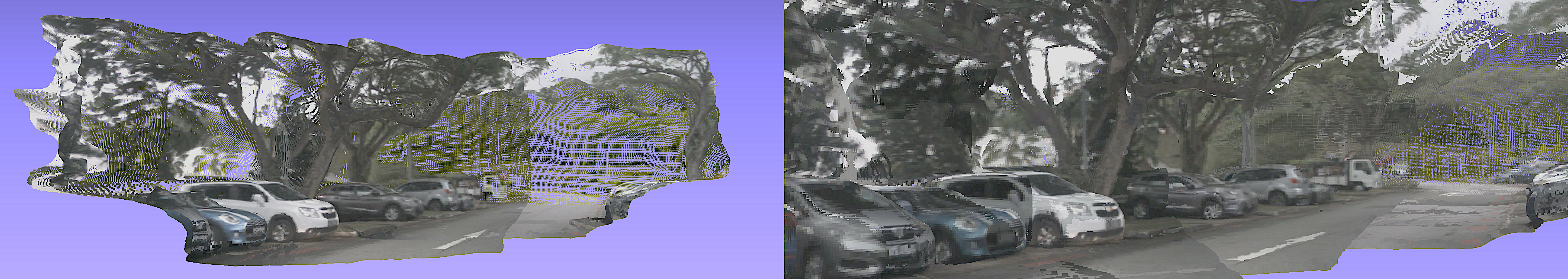}
\vspace*{-0.05in}
\caption{NVS with different frame length: 2(L), 8(R)}
\vspace*{-0.25in}
\label{fig:multiframe}
\end{figure}

\sssec{Attacks:} One concern with novel views is the potential risk of recovering how far away the novel view is from real pose based on the quality of shared rendered images (holes, missing pixels etc). 
It could be resolved by 3D reconstruction over a longer context. For a frame sequence from nuScenes ~\cite{caesar2020nuscenes} dataset, we have observed that the longer sequence we use to reconstruct the 3D point cloud, the fewer missing pixels when we perform NVS at an adjacent viewport, via a plain point cloud reprojection, as shown in ~Fig.\ref{fig:multiframe}.
One could mitigate this by applying random masks over the rendered images (losing some 3D information but protecting any leakage of location) or using inpainting methods \cite{gen3c}. 

\section{Open Questions \& Discussion}
\label{sec:disc}


\sssec{Vision foundation models:} While vision foundation models have revolutionized 3D understanding, key challenges remain: (1) extending robust performance from static to dynamic scenes, (2) obtaining metric-scale rather than relative depth estimates, (3) achieving occlusion-robust view selection, and (4) ensuring resilience against sophisticated adversarial attacks (e.g., physics-based material reflection analysis), (5) extending to other spatial sensors (radar, lidar). Despite these challenges, feedforward models like VGGT show promise for low-latency 3D perception. 

\sssec{Hierarchical data fidelity:} Current cooperative perception systems operate with either all-processed or all-raw data. Given raw data's bandwidth demands, we advocate for hierarchical fidelity with opportunistic switching to maximize data quality while activating privacy preservation when sharing raw data. This requires mechanisms for activating, maintaining, and deactivating \name's privacy methods, alongside strategies to bootstrap open stack adoption.

\sssec{Privacy as a first class citizen for networked vehicles:} Raw data cooperative perception demands privacy-by-design alongside bandwidth, latency, synchronization, sensor interference, and trust considerations. While \name provides high-utility, privacy-safe solutions, it doesn't address all system constraints. Networks may thus contain mixed flows, leading to orchestration and rate control challenges.

\sssec{World privacy:} While roads are public spaces, raw data sharing risks transforming vehicles into mass surveillance systems tracking pedestrians and license plates. Camera data particularly demands privacy safeguards—techniques like face and license plate anonymization before backend upload \cite{tesla} should be incorporated into open stacks to balance raw data benefits with world privacy protection.

\sssec{Implications for off-vehicle world understanding:} Privacy safe raw data processing extends beyond real-time cooperative perception to edge and cloud services like high-resolution map building. While currently dominated by specialized fleets (e.g., Google Street View), \name enables democratized, crowd-sourced mapping accessible to diverse stakeholders (smart city planners, municipalities) at significantly lower cost and higher update frequency.

\section{Conclusion}

Multi-stakeholder groups are currently working to implement vehicle-to-vehicle communication for sharing highly-processed, high-priority safety messages \cite{itsa,landline}. This paper looks even further ahead, exploring the exchange of high-fidelity, raw spatial sensor data between vehicles. For this to be practically adopted, privacy must be a primary design consideration. We've identified several privacy concerns that come with sharing raw data and proposed a comprehensive research agenda to address them. The goal of this paper is to initiate research and international discussions on this topic, aiming for the eventual adoption of raw data networking, similar to what's happening with processed data.




\bibliographystyle{ACM-Reference-Format} 
\bibliography{full}

@article{liu2023echoes,
  author={Debashri Roy and Yuanyuan Li and Tong Jian and Peng Tian and Kaushik Chowdhury and Stratis Ioannidis},
  title={Multi-Modality Sensing and Data Fusion for Multi-Vehicle Detection},
  journal={IEEE Transactions on Multimedia},
  year={2023},
  volume={25},
  pages={2280-2295},
  publisher={Institute of Electrical and Electronics Engineers (IEEE)},
  doi={10.1109/tmm.2022.3145663}
}

@inproceedings{zhang2023robust,
  author={Qingzhao Zhang and Xumiao Zhang and Ruiyang Zhu and Fan Bai and Mohammad Naserian and Z. Morley Mao},
  title={Robust Real-time Multi-vehicle Collaboration on Asynchronous Sensors},
  booktitle={Proceedings of the 29th Annual International Conference on Mobile Computing and Networking},
  year={2023},
  pages={1-15},
  publisher={ACM},
  doi={10.1145/3570361.3613271}
}

@inproceedings{wang2020v2vnet,
  author={Tsun-Hsuan Wang and Sivabalan Manivasagam and Ming Liang and Bin Yang and Wenyuan Zeng and Raquel Urtasun},
  title={V2VNet: Vehicle-to-Vehicle Communication for Joint Perception and Prediction},
  booktitle={ECCV},
  year={2020},
  pages={605-621},
  publisher={Springer International Publishing},
  doi={10.1007/978-3-030-58536-5_36}
}

@inproceedings{xu2022v2x,
  author={Runsheng Xu and Hao Xiang and Zhengzhong Tu and Xin Xia and Ming-Hsuan Yang and Jiaqi Ma},
  title={V2X-ViT: Vehicle-to-Everything Cooperative Perception with Vision Transformer},
  booktitle={ECCV},
  year={2022},
  pages={107-124},
  publisher={Springer Nature Switzerland},
  doi={10.1007/978-3-031-19842-7_7},
  organization={Springer}
}

@article{henry2020pseudonym,
  author={Kevin Henry},
  title={Pseudonym Issuing Strategies for Privacy-Preserving V2X Communication},
  journal={SAE International Journal of Transportation Cybersecurity and Privacy},
  year={2020},
  volume={02},
  number={2},
  pages={131-139},
  publisher={SAE International},
  doi={10.4271/11-02-02-0012}
}

@article{naseer2018robust,
  author={Tayyab Naseer and Wolfram Burgard and Cyrill Stachniss},
  title={Robust Visual Localization Across Seasons},
  journal={IEEE Transactions on Robotics},
  year={2018},
  volume={34},
  number={2},
  pages={289-302},
  publisher={Institute of Electrical and Electronics Engineers (IEEE)},
  doi={10.1109/tro.2017.2788045}
}

@article{toft2020long,
  author={Carl Toft and Will Maddern and Akihiko Torii and Lars Hammarstrand and Erik Stenborg and Daniel Safari and Masatoshi Okutomi and Marc Pollefeys and Josef Sivic and Tomas Pajdla and Fredrik Kahl and Torsten Sattler},
  title={Long-Term Visual Localization Revisited},
  journal={IEEE Transactions on Pattern Analysis and Machine Intelligence},
  year={2022},
  volume={44},
  number={4},
  pages={2074-2088},
  publisher={Institute of Electrical and Electronics Engineers (IEEE)},
  doi={10.1109/tpami.2020.3032010}
}

@inproceedings{wang2025vggt,
  author={Jianyuan Wang and Minghao Chen and Nikita Karaev and Andrea Vedaldi and Christian Rupprecht and David Novotny},
  title={VGGT: Visual Geometry Grounded Transformer},
  booktitle={2025 IEEE/CVF Conference on Computer Vision and Pattern Recognition (CVPR)},
  year={2025},
  pages={5294-5306},
  publisher={IEEE},
  doi={10.1109/cvpr52734.2025.00499}
}

@article{enev2016automobile,
  author={Miro Enev and Alex Takakuwa and Karl Koscher and Tadayoshi Kohno},
  title={Automobile Driver Fingerprinting},
  journal={Proceedings on Privacy Enhancing Technologies},
  year={2016},
  volume={2016},
  number={1},
  pages={34-50},
  publisher={Privacy Enhancing Technologies Symposium Advisory Board},
  doi={10.1515/popets-2015-0029}
}

@inproceedings{das2018deep,
  author={Rajshekhar Das and Akshay Gadre and Shanghang Zhang and Swarun Kumar and Jose M. F. Moura},
  title={A Deep Learning Approach to IoT Authentication},
  booktitle={2018 IEEE International Conference on Communications (ICC)},
  year={2018},
  pages={1-6},
  publisher={IEEE},
  doi={10.1109/icc.2018.8422832}
}

@article{archer2023handbook,
  author={Florian Kerschbaum},
  title={Privacy-Preserving Computation},
  journal={arXiv},
  year={2014},
  pages={41-54},
  publisher={Springer Berlin Heidelberg},
  doi={10.1007/978-3-642-54069-1_3}
}

@misc{kpmg,
  author={Joachim Taiber},
  title={Sharing Data in Automotive Applications},
  year={2020},
  pages={191-204},
  publisher={Springer International Publishing},
  doi={10.1007/978-3-030-35032-1_12}
}

@misc{ftc,
  author={Federal Trade Commission},
  title={Wyoming Federal Court Upholds Law Criminalizing “Unlawful Collection of Resource Data”},
  year={2024},
  publisher={Walter de Gruyter GmbH},
  doi={10.1163/9789004322714_cclc_2016-0142-022}
}

@misc{etsi,
  author={Panagiotis Lytrivis and Angelos Amditis},
  title={Intelligent Transport Systems: Co-Operative Systems (Vehicular Communications)},
  year={2012},
  publisher={InTech},
  doi={10.5772/34970}
}

@BOOK{mikhail2001introduction,
  author={K. B. Atkinson},
  title={Introduction to Modern Photogrammetry.},
  journal={The Photogrammetric Record},
  year={2003},
  volume={18},
  number={104},
  pages={329-330},
  publisher={Wiley},
  doi={10.1046/j.0031-868x.2003.024_01.x}
}

@inproceedings{sumikura2019openvslam,
  author={A. D. Devyatovskaya and N. E. Biryuchkov and T. V. Liakh and K. V. Chaika},
  title={SLAM in Duckietown Simulator Using the OpenVSLAM Framework},
  journal={Vestnik NSU. Series: Information Technologies},
  booktitle={ACM MM},
  year={2022},
  volume={19},
  number={4},
  pages={36-49},
  publisher={Novosibirsk State University (NSU)},
  doi={10.25205/1818-7900-2021-19-4-36-49}
}

@article{mildenhall2021nerf,
  author={Ben Mildenhall and Pratul P. Srinivasan and Matthew Tancik and Jonathan T. Barron and Ravi Ramamoorthi and Ren Ng},
  title={NeRF: Representing Scenes as Neural Radiance Fields for View Synthesis},
  journal={CACM},
  year={2020},
  pages={405-421},
  publisher={Springer International Publishing},
  doi={10.1007/978-3-030-58452-8_24}
}

@inproceedings{fu2024colmap,
  author={Yang Fu and Xiaolong Wang and Sifei Liu and Amey Kulkarni and Jan Kautz and Alexei A. Efros},
  title={COLMAP-Free 3D Gaussian Splatting},
  booktitle={2024 IEEE/CVF Conference on Computer Vision and Pattern Recognition (CVPR)},
  year={2024},
  pages={20796-20805},
  publisher={IEEE},
  doi={10.1109/cvpr52733.2024.01965}
}

@inproceedings{bansal2020pointillism,
  author={Danfei Xu and Dragomir Anguelov and Ashesh Jain},
  title={PointFusion: Deep Sensor Fusion for 3D Bounding Box Estimation},
  booktitle={2018 IEEE/CVF Conference on Computer Vision and Pattern Recognition},
  year={2018},
  pages={244-253},
  publisher={IEEE},
  doi={10.1109/cvpr.2018.00033}
}

@inproceedings{godard2019digging,
  author={Clement Godard and Oisin Mac Aodha and Michael Firman and Gabriel Brostow},
  title={Digging Into Self-Supervised Monocular Depth Estimation},
  booktitle={2019 IEEE/CVF International Conference on Computer Vision (ICCV)},
  year={2019},
  publisher={IEEE},
  doi={10.1109/iccv.2019.00393}
}

@misc{tesla,
  author={Alan Tang},
  title={Privacy Notice},
  year={2022},
  pages={165-176},
  publisher={CRC Press},
  doi={10.1201/9781003225089-19}
}

@misc{landline,
  author={Peter Suber},
  title={Open Everything launches today},
  year={2024},
  publisher={Front Matter},
  doi={10.63485/r5gtq-f9q36}
}

@misc{itsa,
  author={ITS America},
  title={LTE-V2X Requirements and Deployment Profile for Aftermarket V2X Devices},
  year={2023},
  publisher={SAE International},
  doi={10.4271/j3315_202510}
}

@inproceedings{khan2024vrf,
  author={Kaleem Nawaz Khan and Ali Khalid and Yash Turkar and Karthik Dantu and Fawad Ahmad},
  title={VRF: Vehicle Road-side Point Cloud Fusion},
  booktitle={Proceedings of the 22nd Annual International Conference on Mobile Systems, Applications and Services},
  year={2024},
  pages={547-560},
  publisher={ACM},
  doi={10.1145/3643832.3661874}
}

@inproceedings{xu2022opencood,
  author={Neddly Maxime},
  title={Carla},
  booktitle={ACM SIGGRAPH 98 Conference abstracts and applications},
  year={1998},
  pages={159},
  publisher={ACM},
  doi={10.1145/280953.284834}
}

@inproceedings{colmap,
  author={Johannes L. Schonberger and Jan-Michael Frahm},
  title={Structure-from-Motion Revisited},
  booktitle={2016 IEEE Conference on Computer Vision and Pattern Recognition (CVPR)},
  year={2016},
  publisher={IEEE},
  doi={10.1109/cvpr.2016.445}
}

@inproceedings{gen3c,
  author={Xuanchi Ren and Tianchang Shen and Jiahui Huang and Huan Ling and Yifan Lu and Merlin Nimier-David and Thomas Müller and Alexander Keller and Sanja Fidler and Jun Gao},
  title={Gen3C: 3D-Informed World-Consistent Video Generation with Precise Camera Control},
  booktitle={2025 IEEE/CVF Conference on Computer Vision and Pattern Recognition (CVPR)},
  year={2025},
  pages={6121-6132},
  publisher={IEEE},
  doi={10.1109/cvpr52734.2025.00574}
}

@inproceedings{shenoy2022rf,
  author={Jingyang Hu and Hongbo Jiang and Siyu Chen and Qibo Zhang and Zhu Xiao and Daibo Liu and Jiangchuan Liu and Bo Li},
  title={WiShield: Privacy Against Wi-Fi Human Tracking},
  journal={IEEE Journal on Selected Areas in Communications},
  booktitle={SIGCOMM},
  year={2024},
  volume={42},
  number={10},
  pages={2970-2984},
  publisher={Institute of Electrical and Electronics Engineers (IEEE)},
  doi={10.1109/jsac.2024.3414597}
}

@inproceedings{dust3r,
  author={Shuzhe Wang and Vincent Leroy and Yohann Cabon and Boris Chidlovskii and Jerome Revaud},
  title={DUSt3R: Geometric 3D Vision Made Easy},
  booktitle={2024 IEEE/CVF Conference on Computer Vision and Pattern Recognition (CVPR)},
  year={2024},
  pages={20697-20709},
  publisher={IEEE},
  doi={10.1109/cvpr52733.2024.01956}
}

@inproceedings{what2comm,
  author={Kun Yang and Dingkang Yang and Jingyu Zhang and Hanqi Wang and Peng Sun and Liang Song},
  title={What2comm: Towards Communication-efficient Collaborative Perception via Feature Decoupling},
  booktitle={Proceedings of the 31st ACM International Conference on Multimedia},
  year={2023},
  pages={7686-7695},
  publisher={ACM},
  doi={10.1145/3581783.3611699}
}

@inproceedings{where2comm,
  author={Yue Hu and Juntong Peng and Sifei Liu and Junhao Ge and Si Liu and Siheng Chen},
  title={Communication-Efficient Collaborative Perception via Information Filling with Codebook},
  journal={NeurIPS},
  booktitle={2024 IEEE/CVF Conference on Computer Vision and Pattern Recognition (CVPR)},
  year={2024},
  pages={15481-15490},
  publisher={IEEE},
  doi={10.1109/cvpr52733.2024.01466}
}

@inproceedings{how2comm,
  author={Seungjun Hyeon and Minho Park and Dong-oh Kang},
  title={Network-Based Message Selection for Efficient Communication of Multi-Agent Collaboration},
  journal={NeurIPS},
  booktitle={2024 15th International Conference on Information and Communication Technology Convergence (ICTC)},
  year={2024},
  pages={2188-2189},
  publisher={IEEE},
  doi={10.1109/ictc62082.2024.10826897}
}

@article{droid,
  author={Kazuki Adachi and Yoshitaka Hara and Sousuke Nakamura},
  title={Simulation Evaluation of Monocular Visual SLAM: ORB-SLAM3, DROID-SLAM, DPVO, and DPV-SLAM},
  journal={NeurIPS},
  booktitle={2025 IEEE International Conference on Multisensor Fusion and Integration for Intelligent Systems (MFI)},
  year={2025},
  pages={1-5},
  publisher={IEEE},
  doi={10.1109/mfi67357.2025.11259365}
}

@article{3dgs,
  author={Bernhard Kerbl and Georgios Kopanas and Thomas Leimkuehler and George Drettakis},
  title={3D Gaussian Splatting for Real-Time Radiance Field Rendering},
  journal={ACM Transactions on Graphics},
  year={2023},
  volume={42},
  number={4},
  pages={1-14},
  publisher={Association for Computing Machinery (ACM)},
  doi={10.1145/3592433}
}

@article{keetha2025mapanything,
  author={Noah Stier and Anurag Ranjan and Alex Colburn and Yajie Yan and Liang Yang and Fangchang Ma and Baptiste Angles},
  title={FineRecon: Depth-aware Feed-forward Network for Detailed 3D Reconstruction},
  journal={arXiv},
  booktitle={2023 IEEE/CVF International Conference on Computer Vision (ICCV)},
  year={2023},
  pages={18377-18386},
  publisher={IEEE},
  doi={10.1109/iccv51070.2023.01689}
}

@inproceedings{caesar2020nuscenes,
  author={Holger Caesar and Varun Bankiti and Alex H. Lang and Sourabh Vora and Venice Erin Liong and Qiang Xu and Anush Krishnan and Yu Pan and Giancarlo Baldan and Oscar Beijbom},
  title={nuScenes: A Multimodal Dataset for Autonomous Driving},
  booktitle={2020 IEEE/CVF Conference on Computer Vision and Pattern Recognition (CVPR)},
  year={2020},
  publisher={IEEE},
  doi={10.1109/cvpr42600.2020.01164}
}

@inproceedings{recap,
  author={Christina Suyong Shin and Weiwu Pang and Chuan Li and Fan Bai and Fawad Ahmad and Jeongyeup Paek and Ramesh Govindan},
  title={RECAP: 3D Traffic Reconstruction},
  booktitle={Proceedings of the 30th Annual International Conference on Mobile Computing and Networking},
  year={2024},
  pages={1252-1267},
  publisher={ACM},
  doi={10.1145/3636534.3690691}
}

\end{document}